\documentstyle[aps,epsf,preprint]{revtex}
\tightenlines
\begin{document}
\draft
\title{Charged Kaon and Pion Production at Midrapidity
in Proton Nucleus and Sulphur Nucleus Collisions}
\author {H.~B\o ggild$^{1}$, J.~Boissevain$^{2}$,
J.~Dodd$^{3}$, S.~Esumi$^{4,a}$, 
C.W.~Fabjan$^{5}$, D.E.~Fields$^{2,b}$, A.~Franz$^{5,c}$, 
K.H.~Hansen$^{1,\dagger}$,
T.J.~Humanic$^{6}$, B.V.~Jacak$^{7}$, 
H.~Kalechofsky$^{8}$, Y.Y.~Lee$^{8}$, M.~Leltchouk$^{3}$,
 B.~L{\"o}rstad$^{9}$, 
N.~Maeda$^{4,d}$, 
A.~Miyabayashi$^{9}$, M.~Murray$^{10}$, 
S.~Nishimura$^{4,e}$,  S.U.~Pandey$^{6,f}$, F.~Piuz$^{5}$, 
V.~Polychronakos$^{11}$, M.~Potekhin$^{3}$,
G.~Poulard$^{5}$, A.~Sakaguchi$^{4,g}$, 
M.~Sarabura$^{2}$, J.~Simon-Gillo$^{2}$, J.~Schmidt-S\o rensen$^{9}$,  
W.~Sondheim$^{2}$, 
T.~Sugitate$^{4}$, J.P.~Sullivan$^{2}$, Y.~Sumi$^{4}$, 
H.~van~Hecke$^{2}$, W.J.~Willis$^{3}$ and K.~Wolf$^{10,\dagger}$\\
\center{ (The NA44 Collaboration) nucl-ex/9808002\\}}

\address{
$^{1}$ Niels Bohr Institute, DK-2100 Copenhagen, Denmark.\\
$^{2}$ Los Alamos National Laboratory, Los Alamos, NM 87545\\
$^{3}$ Columbia University, New York, NY 10027\\
$^{4}$ Hiroshima University, Higashi-Hiroshima 739, Japan.\\
$^{5}$ CERN, CH-1211 Geneva 23, Switzerland.\\
$^{6}$ Ohio State University, Columbus,OH 43210\\
$^{7}$ State University of New York Stony Brook, Stony Brook, NY 11794\\
$^{8}$ University of Pittsburgh, Pittsburgh, PA 15260\\
$^{9}$ University of Lund, S-22362 Lund, Sweden.\\
$^{10}$ Texas A\&M University, College Station, TX 77843-3366\\
$^{11}$ Brookhaven National Laboratory, Upton, NY 11973\\
$^a$ Now at Heidelberg University, Heidelberg, D-69120, Germany\\
$^b$ Now at University of New Mexico, Albuquerque, NM 87185\\
$^c$ Now at Brookhaven National Laboratory, Upton, NY 11973\\
$^d$ Now at Florida State University, Tallahassee, FL 32306\\
$^e$ Now at University of Tsukuba, Ibaraki 305, Japan\\
$^f$ Now at Wayne State University, Detroit, MI 48201\\
$^g$ Now at Osaka University, Osaka 560, Japan\\
$^\dagger$ deceased}
 
\maketitle
\newpage
\begin{abstract}
The NA44 collaboration has  measured charged kaon and pion distributions
 at midrapidity in
sulphur and proton collisions with nuclear targets at  200
and 450 GeV/{\it c} per nucleon, respectively.
The inverse slopes of 
kaons are larger than those of pions. 
 The difference in the inverse slopes of pions, kaons and
protons, all measured in our spectrometer, increases with system size and 
is consistent with the buildup of collective flow for larger systems.
The target dependence of both the yields and inverse slopes is stronger for
the sulphur beam suggesting the increased importance of secondary 
rescattering
for $SA$ reactions. 
The rapidity density, dN/dy, of both $K^+$ and $K^-$ increases more
rapidly with system size than for $\pi^+$ in a similar rapidity region.
This trend continues with increasing centrality, and according to RQMD, it is caused by
secondary reactions between mesons  and baryons. 
 The $K^-/K^+$ ratio falls
with increasing system size but more slowly than the $\bar{p}/p$ ratio.
The $\pi^-/\pi+$ ratio is close to unity for all systems.
From $pBe$ to $SPb$  the 
$K^+/p$ ratio decreases while $K^-/\bar{p}$ increases and 
$\sqrt{\frac{K^+\cdot K^-}{p \cdot \bar{p}}} $ stays constant.
These data suggest that as larger nuclei collide, the resulting system
 has a larger transverse expansion,  baryon density and an increasing 
fraction of strange quarks.

\end{abstract}


\section{Introduction}

Nucleus-nucleus collisions at ultrarelativistic energies
create hadronic matter at high energy density. 
Enhanced production of particles containing a strange quark, such as kaons,
 may indicate formation of
a state of matter in which the quarks and gluons are deconfined
\cite{Hei84a,Koc88a,Ell89a,Lee88a}. This is because while the production 
of s-quarks is generally suppressed in hadronic collisions, they are light
enough to be abundantly produced at temperatures above the deconfinement
 phase transition,
and rapidly reach chemical equilibrium via
gluon-gluon fusion (see \cite{MULL96A} for a review).
Alternatively there may be a duality
of quark matter and ``conventional" hadronic explanations for the
strangeness enhancement observed in recent data \cite{SOR98A}.
 The $K^-$ and $K^+$
yields together provide a sensitive probe of the
spacetime evolution of heavy-ion reactions. 
Since $K^-$s have a large annihilation cross-section with neutrons, their
 yield is sensitive to the baryon density. 
The $K^-$ and $K^+$ distributions may
also hint at the degree of thermalization achieved;
and their transverse mass spectra 
allow detailed study of rescattering and collective expansion effects.
The pions are less sensitive to collective expansion because of their
small mass. However because they are the most numerous of the produced
particles they give information on the total entropy produced in the
collision. 
We present charged kaon and pion measurements using the NA44
spectrometer
from $pBe$ collisions (to approximate $pp$), p-nucleus and 
nucleus-nucleus interactions. This allows a systematic study 
as a function of the size of the central
region and different conditions in the surrounding hadronic matter.

The distributions of kaons at midrapidity provide a sensitive probe 
of the collision dynamics and constrain the assumptions of event generators.
We compare our data to 
the RQMD model, version 1.08 \cite{Sor89a,Sor90a}. RQMD is a 
microscopic phase space approach to modeling
relativistic heavy ion collisions 
based on resonance and string
excitation and fragmentation with subsequent hadronic collisions.
RQMD also contains ``ropes" formed by 
overlapping strings, 
 whose sources are color octet charge states \cite{Sor92a}. These
increase the number of strange antibaryons and antiprotons but 
have little  effect on kaon yields or slopes.

\section{Experiment}

\begin{figure}
     \epsfxsize=16cm
     \epsffile{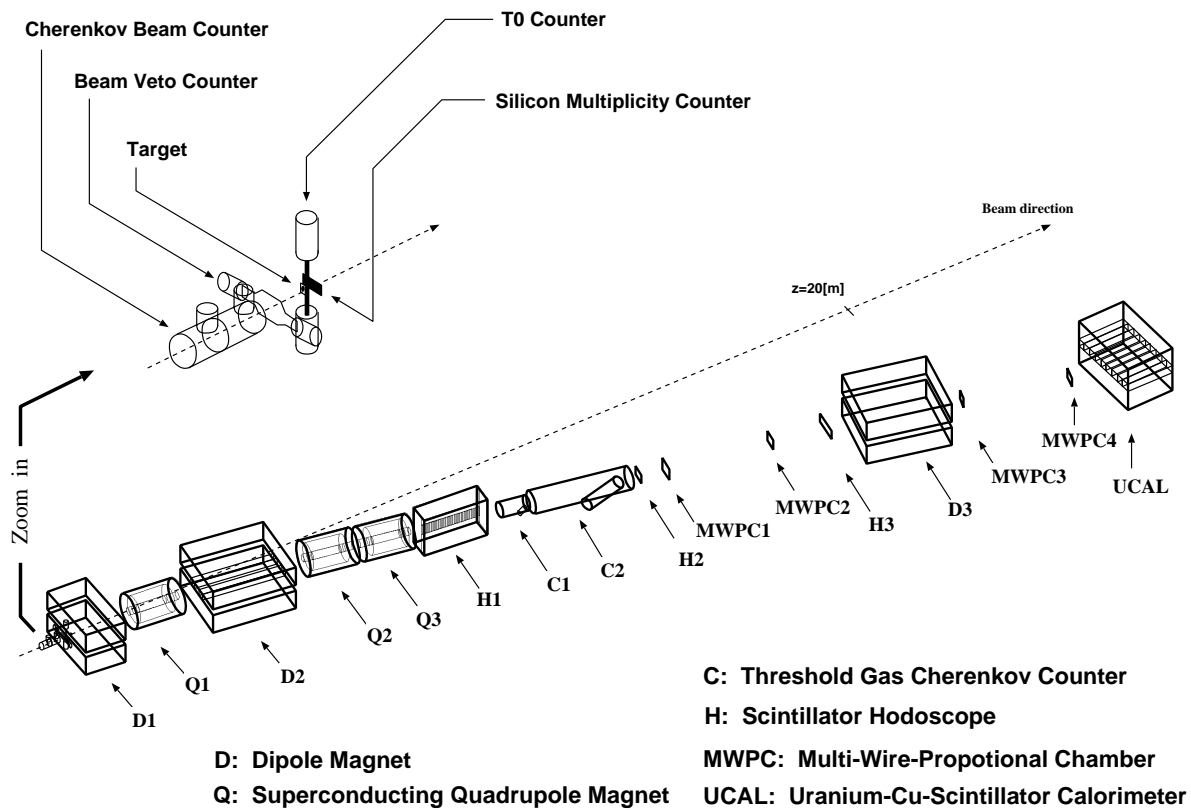}
    \caption{The NA44 experimental set-up.}
    \label{fg:setup}
  \end{figure}

The NA44 experiment is shown in Fig.  \ref{fg:setup}.
Three conventional dipole magnets (D1, D2, and D3) and three 
superconducting quadrupoles 
(Q1, Q2, and Q3) analyze the momentum and create a magnified 
image of the target in the spectrometer. The magnets focus particles 
from the target onto the first hodoscope (H1) such that
the horizontal position along the hodoscope gives the total momentum.
Two other hodoscopes (H2, H3) measure the angle of the 
track.
The hodoscopes also measure the time-of-flight with a resolution
of approximately 100~ps. Particle identification relies 
primarily upon the third hodoscope.

\begin{figure}
    \begin{center}\mbox{  
        \epsfxsize=12cm
        \epsffile{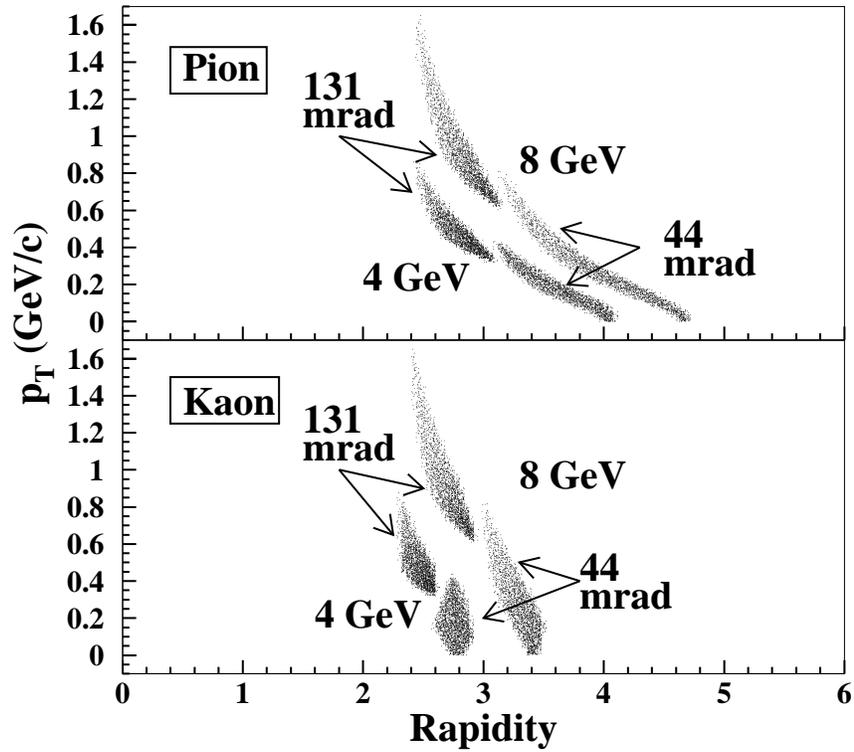}}\end{center}
    \caption{The acceptance of charged pions and kaons 
    in y and $p_{T}$.
 The $\phi$ acceptance (not shown) decreases from
 $2\pi$ at $p_{T}$~=~0.0~GeV/{\it c} to 0.1 at $p_{T}$~=~0.8~GeV/{\it c}.}
    \label{fg:ac_kpi}
\end{figure}

The momentum acceptance 
is $\pm$ 20\% of the nominal momentum setting. 
The angular coverage, with respect to the beam, is approximately 
$\pm$ 5 mrad vertically
and -5 to +78 mrad 
in the
horizontal plane 
when the spectrometer is at 44mrad with respect to the beam.
When the spectrometer is rotated to 131mrad, the horizontal coverage is 77 to 165 mrad. 
Only particles of a fixed charge sign are detected in a given 
spectrometer setting.
Four settings are used to cover the midrapidity 
region in the $p_T$ range 0 to 1.6 GeV/{\it c}. 
 Fig.  \ref{fg:ac_kpi}
shows the acceptance of the spectrometer for pions and 
kaons in the $y-p_T$
plane for the 4 and 8GeV/{\it c} momentum settings when the 
spectrometer axis is at  44 and 131 mrad with respect to the beam.

\begin{figure}
    \begin{center}\mbox{  
        \epsfxsize=12cm
        \epsffile{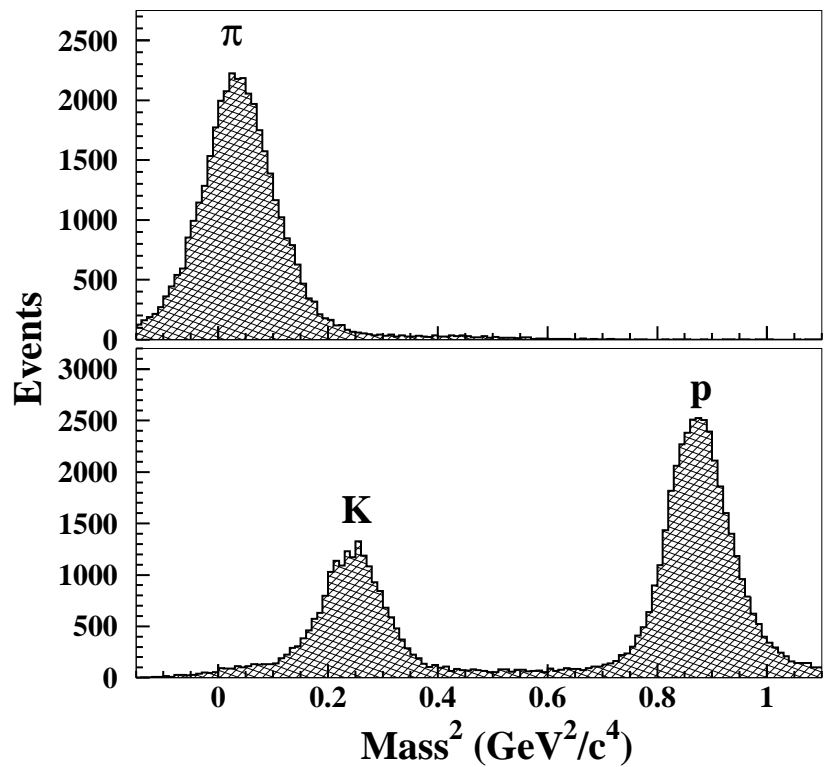}}\end{center}
    \caption{Mass-squared distribution from  Hodoscope 3 with Cerenkov 1 in 
coincidence for pions (top) and veto mode for kaons (bottom).}
    \label{fg:pidkpi}
\end{figure}

 Two Cherenkov counters differentiate kaons, protons
and pions and reject electrons. An appropriate combination 
of C1 and C2 was used for each
spectrometer setting to trigger on pion or kaon/proton events.
Particles are identified by their mass-squared, 
constructed from the track momentum and time-of-flight,
in addition to the Cherenkov information.
This is shown in Fig.~\ref{fg:pidkpi} for the 4GeV/{\it c} setting.
Pions are clearly visible after kaons and protons have been rejected
by requiring a signal in the first Cherenkov counter.
Kaons and protons are clearly separated after pions have been rejected
using the same Cherenkov. 
 After mass-squared and Cherenkov cuts the residual 
contamination of the data is less than 3\%.

A scintillator (T$_0$) is used to trigger on central events in
sulphur-nucleus collisions by requiring
a large pulse height (high charged particle multiplicity). 
The pseudorapidity coverage of T$_0$ is roughly 1.3 to 3.5.
For proton beams, T$_0$ provides
the interaction trigger by requiring that at least one charged particle hit
the scintillator.
 More details about the spectrometer are 
available in \cite{NA44mt,NA44pmp}.

\section{Data Analysis}

 The  data samples after particle identification 
and quality cuts are shown in Table~\ref{tb:data}.
Also shown for each data set is the target 
thickness and the centrality, expressed as a fraction of 
the  total inelastic cross-section. 
In order to construct the invariant cross-section,
the raw distributions are corrected using 
a Monte Carlo 
simulation of the detector response. 
 Simulated tracks are passed 
through the full analysis software chain and used to correct the data 
for geometrical acceptance, reconstruction efficiency,
particle decay in flight and momentum 
resolution. Particles are generated uniformly in rapidity and 
according to an exponential
distribution in transverse mass, $m_T=\sqrt{p_T^2+m^2}$,
with the coefficient of the exponent determined
iteratively from the data.

The absolute normalization of each spectrum is calculated
using the number of beam particles,
the target thickness, the fraction of interactions satisfying 
the trigger, and the measured live time of the data-acquisition
system. For the $SA$ data, the centrality
selection is determined by comparing the pulse height
distribution in the T$_0$ counter 
for central and unbiased beam triggers.
For $pA$ systems, the 
fraction of inelastic collisions
producing at least one hit in the interaction (T$_0$)
counter is modeled with the event
 generators RQMD \cite{Sor89a} and FRITIOF \cite{FRIT87,FRIT92}.
The resulting centrality fractions are indicated in 
Table~\ref{tb:data}. 

The proton-nucleus data are  corrected for non-target background.
The largest corrections are 15\%, 12\% and 12\% for $K^+$, $K^-$ and $\pi^+$
respectively,
for $pBe$ collisions. This correction does not affect the shape 
of the distribution, only the absolute cross-section.
No correction is needed for the central nucleus-nucleus data.
The cross-sections are also corrected for
the particle identification and trigger inefficiencies.

The invariant cross-sections, measured in the NA44 acceptance, are
generally exponentials in transverse mass. 
This allows us to characterize the distributions by two
numbers: the inverse slope, and the total particle yield.
A detailed description of the estimated 
systematic errors on particle yields and inverse slopes 
is given in \cite{NA44pmp}. These were derived from checks 
on the momentum scale, acceptance corrections, trigger and 
lifetime efficiencies, see Table~\ref{tb:errors}.
\section{Results}
 
\begin{figure}
    \begin{center}\mbox{  
        \epsfxsize=12cm
        \epsffile{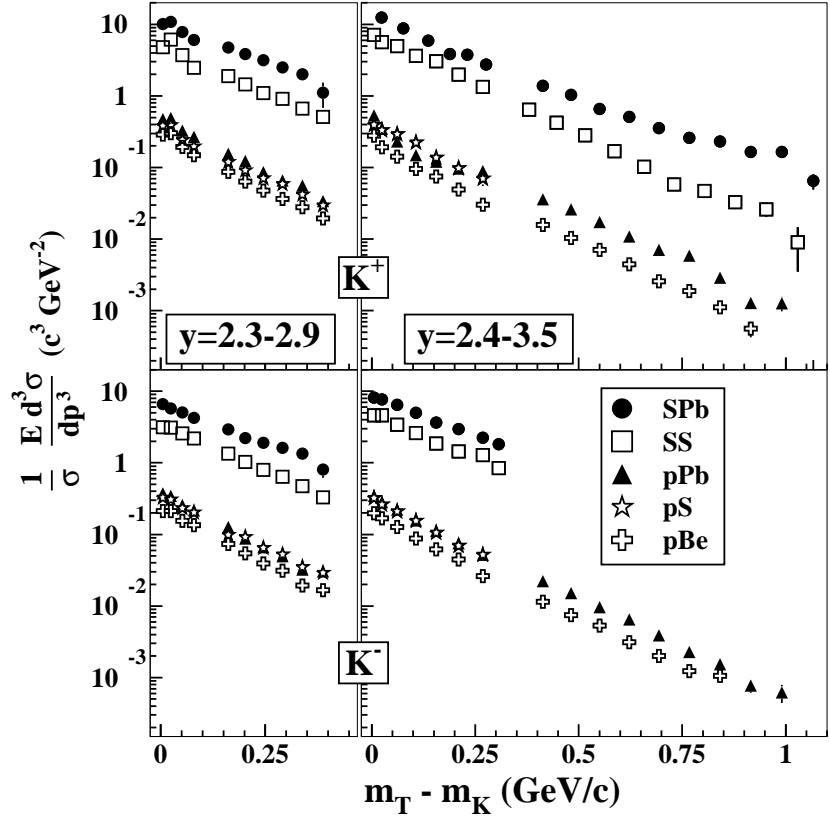}}\end{center}
    \caption{Invariant cross-sections as a function of $m_T - m_K$ for
$K^+$ and $K^-$.}
    \label{fg:kpm48}
\end{figure}

\begin{figure}
    \begin{center}\mbox{  
        \epsfxsize=12cm
        \epsffile{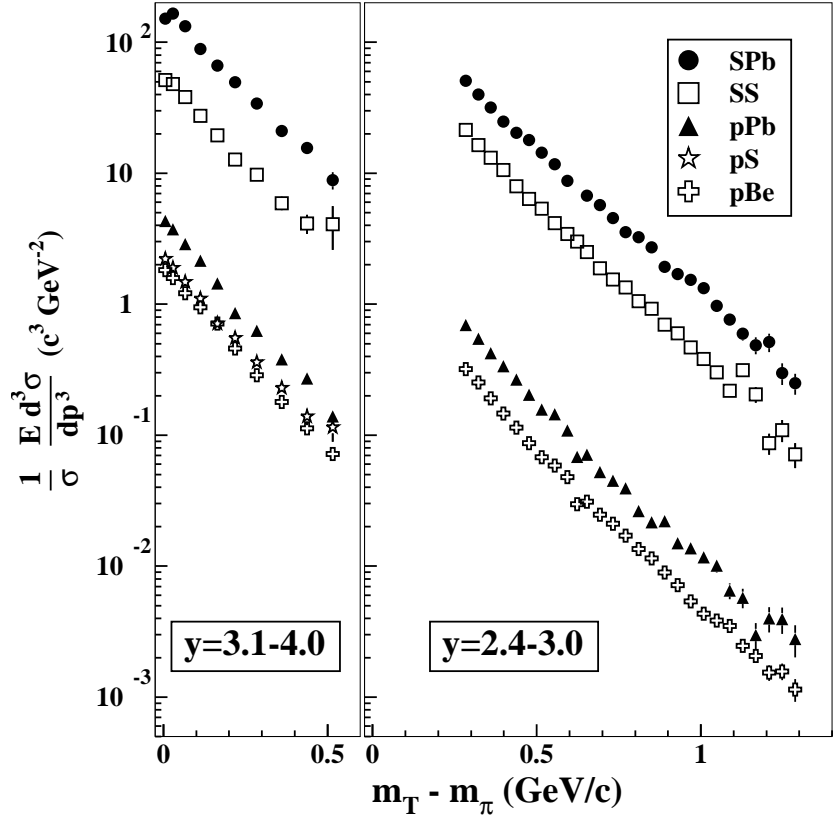}}\end{center}
    \caption{Invariant cross-sections as a function of $m_T - m_\pi$ for
$\pi^+$. These spectra were formed by merging the 4 and 8~GeV/{\it c}
spectra from the 44mrad and 131 mrad settings.}
    \label{fg:pi48lohi}
\end{figure}

 The invariant cross-sections for kaons and pions from
$pBe$, $pS$, $pPb$, $SS$,  and $SPb$ collisions are shown
in Figs.~\ref{fg:kpm48} and ~\ref{fg:pi48lohi} as a function of 
$m_T-m$. The transverse
mass distributions are generally described by exponentials in the 
region of the NA44 acceptance, that is:-
\begin{equation}
 \frac{1}{\sigma} \frac{E d^3\sigma}{d^3p} = Ce^{-(m_T-m)/T},
\label{eq:mt}
\end{equation}
where $C$ is a constant and  $T$ is the inverse logarithmic slope.
The pion spectra are not exponential for all $m_T$ however as 
discussed in Section~\ref{sec-disc}.

\begin{figure}
  \begin{center}\mbox{  
  \epsfxsize=12cm
  \epsffile{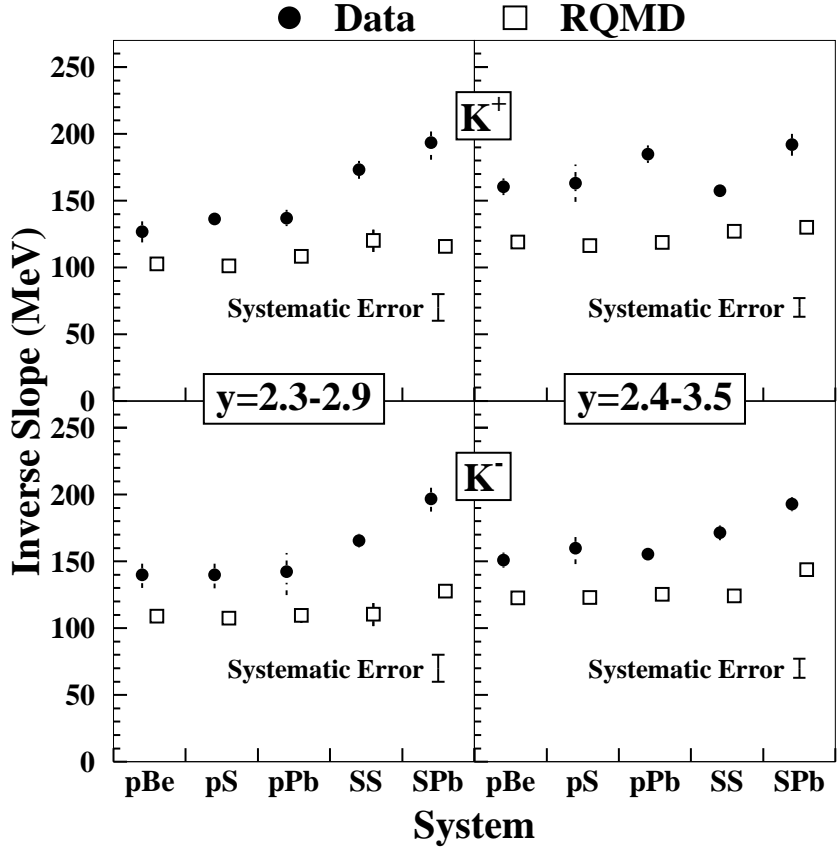}}\end{center}
  \caption{Inverse slopes of the transverse mass distributions for 
    kaons from each system for
    data and RQMD. 
    The global systematic errors common
    to all systems (Table~\protect{\ref{tb:errors}})
    are shown by bars near
the bottom right-hand corner of each plot.
For RQMD, there is no significant
difference in the inverse slopes if rope formation is turned off.
The shape of the RQMD spectra is concave in the NA44 acceptance, resulting
in slopes that are smaller than the data. Further study is necessary to 
understand why this is so.}
  \label{fg:ktvrqmd}
\end{figure}

\begin{figure}
  \begin{center}\mbox{  
  \epsfxsize=12cm
  \epsffile{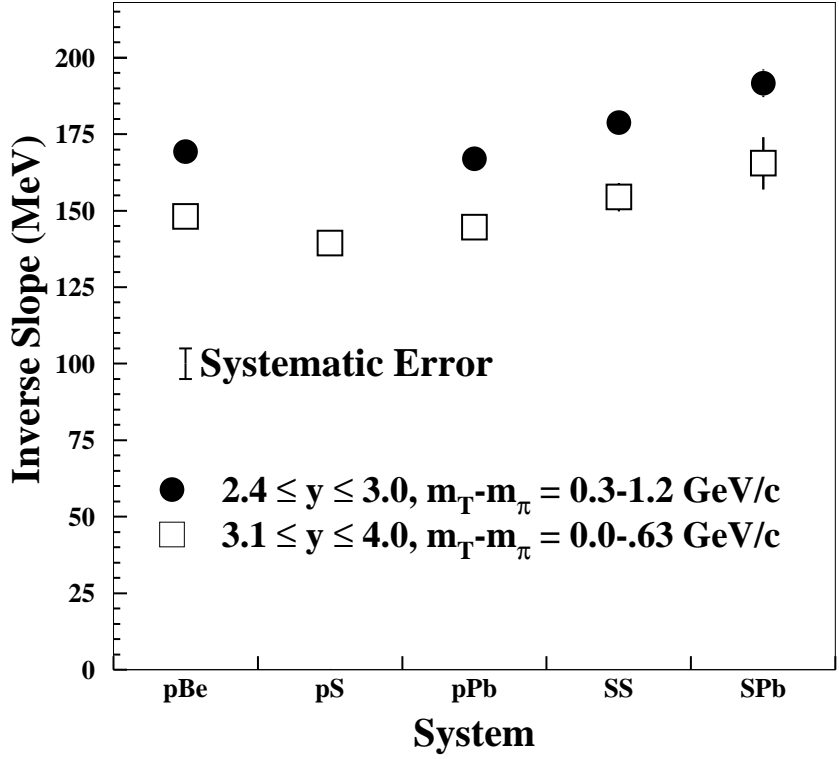}}\end{center}
  \caption{Inverse slopes of the transverse mass distributions for 
pions from each system for our two rapidity intervals.
    The global systematic error common
    to all systems (Table~\protect{\ref{tb:errors}}) is 
    shown by the bars near the  left-hand side of the plot.
 The inverse slopes of $\pi^-$ and $\pi^+$ are equal within errors.}
  \label{fg:pislope}
\end{figure}

The inverse slope parameters obtained by fitting the 
spectra to Eq.~\ref{eq:mt} 
are given in Table~\ref{tb:slope}
and plotted in Figs.~\ref{fg:ktvrqmd} and ~\ref{fg:pislope}. 
The inverse slope parameters for both $K^+$ and $K^-$ increase with
system size.
If the $SPb$ sample is split into two centrality bins,
$11-6\%$ and $6-0\%,$ the inverse slopes 
for both $K^+$ and $K^-$ show no centrality dependence.
The inverse slopes of kaons from RQMD are smaller 
than the data. 
The shape of the RQMD spectra for kaons is concave in the NA44 
acceptance, resulting in slopes that are smaller than the data.
 Further study is necessary to 
understand why this is so. 
The inverse slopes of pions show a weaker
system dependence than kaons but increase with $m_T$.

\begin{figure}
  \begin{center}\mbox{  
        \epsfxsize=12cm
        \epsffile{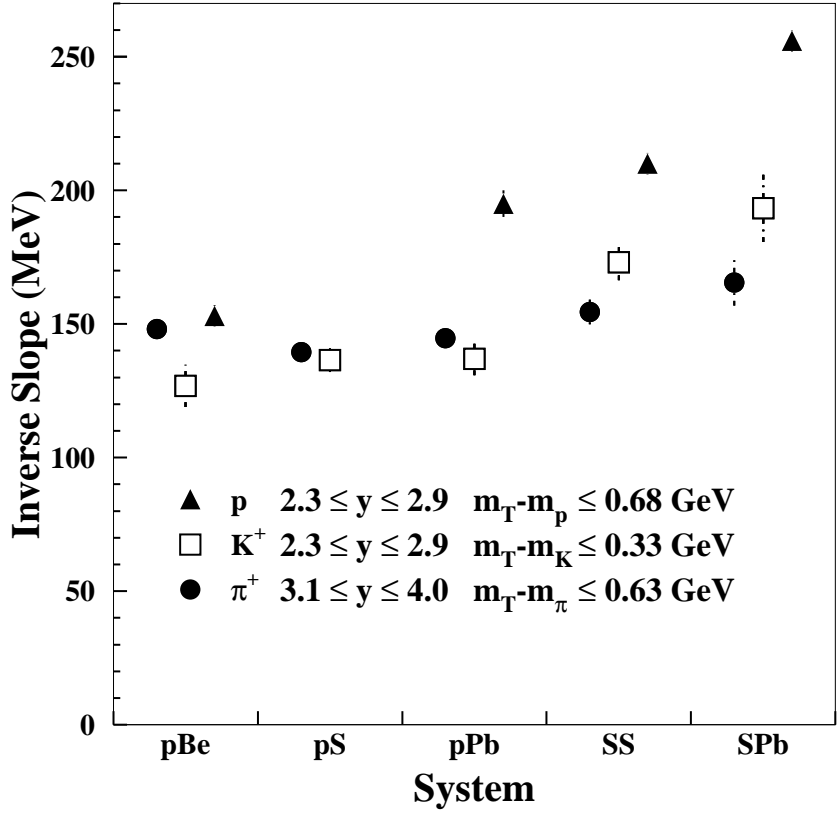}}\end{center}
    \caption{The inverse slopes of pions kaons and protons 
 vs system, the errors are statistical. Correcting for weak decays would
increase the proton inverse slopes by $2\pm1\%$ for $pBe$ and $10\pm2\%$ for
 $SPb$, see ~\protect{\cite{NA44pmp}}.}  
    \label{fg:tpikp}
  \end{figure}

The inverse slopes of our $\pi^+$, $K^+$ and proton \cite{NA44pmp} spectra
 are shown in Fig.~\ref{fg:tpikp}  versus system.
The inverse slopes  of pions increase only slowly with system size, while
those of kaons and protons increase more rapidly. The difference in slopes 
increases with system size. This is consistent with transverse flow, which
should be stronger for larger systems.

The rapidity density, dN/dy,
was calculated by integration of the normalized $m_T - m$ 
distributions, corrected as described above. The fitted slope was 
used to extrapolate to high $m_T$, beyond the region of measurement. 
The statistical error on this extrapolation
was calculated using the full error matrix from the fit of 
Eq.~\ref{eq:mt} to the $m_T$ spectrum.

\begin{figure}
  \begin{center}\mbox{  
  \epsfxsize=12cm
  \epsffile{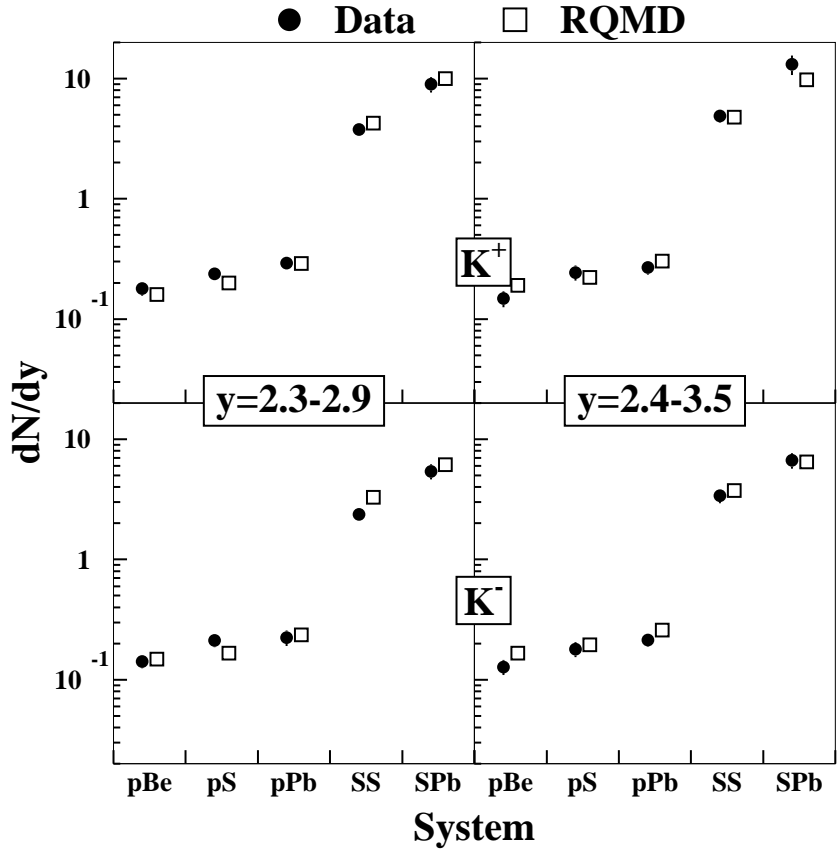}}\end{center}
  \caption{The kaon dN/dy from each system for
    data and RQMD. Statistical and systematic errors for the data
    are added in quadrature.}
  \label{fg:knvrqmd}
\end{figure}

The $K^+$ and $K^-$ dN/dy values from data and RQMD are
plotted in Fig.~\ref{fg:knvrqmd} and
the data (including the pion dN/dy values) are given in Table~\ref{tb:dndy}. 
The uncertainties in dN/dy introduced by the uncertainty
in the slope of the $m_T$ spectrum, the Cerenkov trigger efficiency and the 
combining of the spectra from the two settings 
 are included in the 
errors.
The number of kaons
found at midrapidity increases rapidly with the size of the
colliding nuclei. Comparing ($pPb/pS$) with ($SPb/SS$)
shows that the target dependence is stronger in nucleus-nucleus
collisions than in p-nucleus collisions.

\begin{figure}
    \begin{center}\mbox{  
        \epsfxsize=12cm
        \epsffile{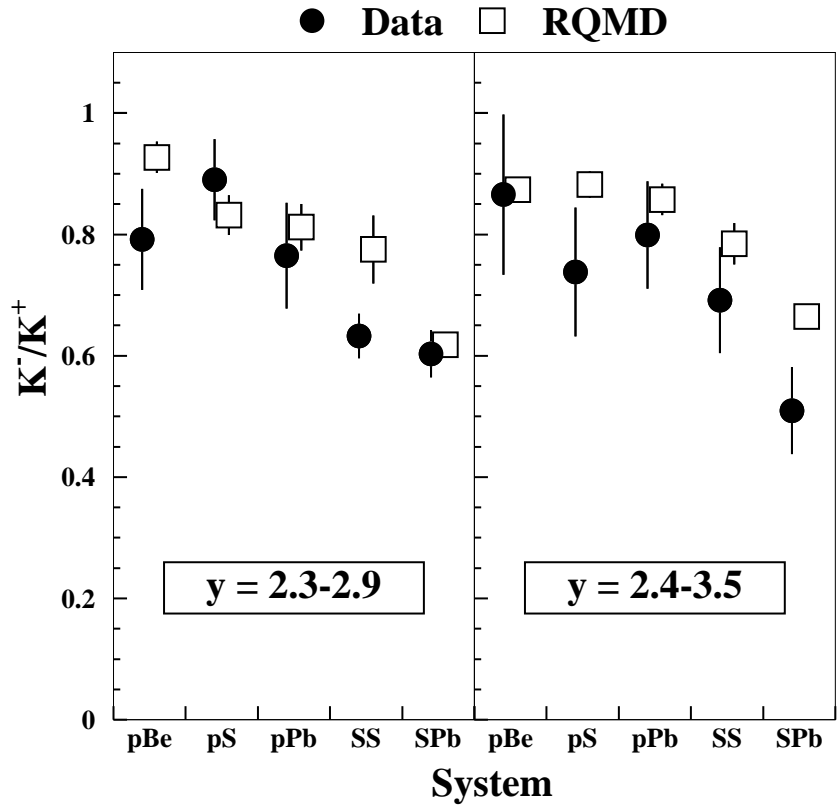}}\end{center}
    \caption{The $K^-/K^+$ ratio vs system for data and RQMD.
 Systematic and statistical errors have been added in quadrature.}
    \label{fg:kratvqm}
\end{figure}

Fig.~\ref{fg:kratvqm} shows the ratio of dN/dy for 
$K^-$ and $K^+$ for the various target-projectile systems studied.
Note that the systematic errors listed in Table~\ref{tb:errors} cancel 
in this ratio. The ratio for $pBe$ agrees with $K^-/K^+$ from
pp collisions measured at the ISR \cite{GUE76A,GUE77A}.
 The ratio drops slowly with system size 
but this is not explained by the lower
$\sqrt{s}$ of the $SA$ data. Only a $3\%$ change in $K^-/K^+$ was 
observed for $pp$ collisions when changing $\sqrt{s}$ from 
31 to ~23GeV/{\it c} \cite{GUE76A}.
The ratio also decreases for more central collisions.
Note the $\pi^-/\pi^+$ ratio is $1.0 \pm .15$ for all systems.
RQMD reproduces the $K^-/K^+$ ratio. 

\begin{figure}
    \begin{center}\mbox{  
        \epsfxsize=12cm
        \epsffile{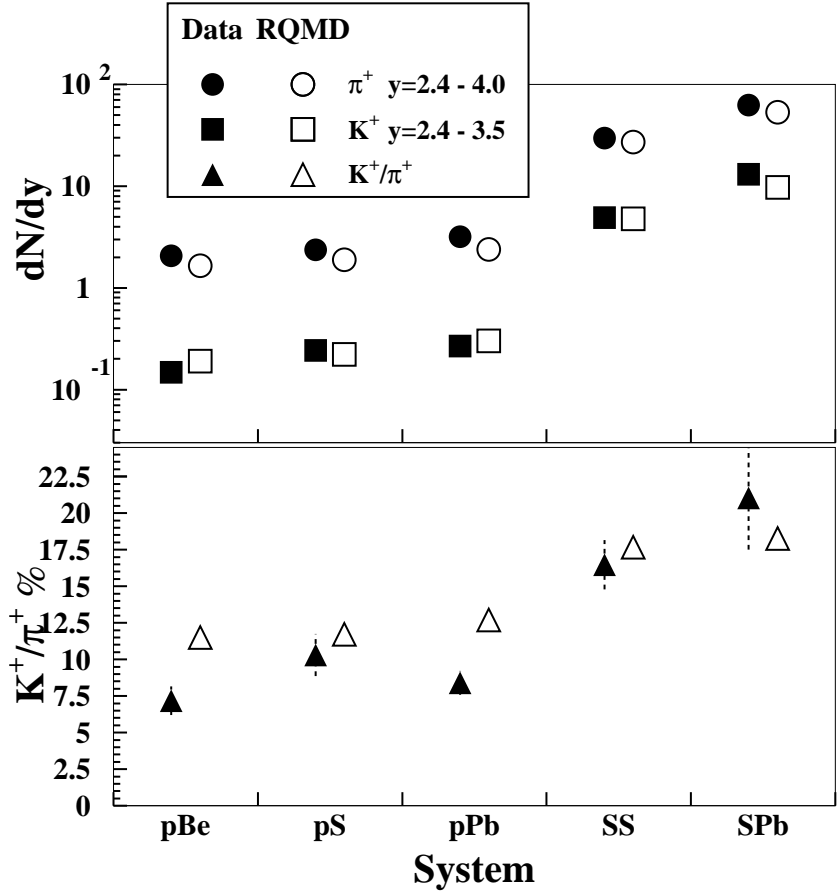}}\end{center}
    \caption{The dN/dy of positive pions and kaons vs system for data
and RQMD.
 Systematic and statistical errors have been added in quadrature.}
    \label{fg:kpidndy}
  \end{figure}

Fig.~\ref{fg:kpidndy} shows the yield of positive pions and kaons 
and their ratio for $y\approx 3$ vs system.
For pions, like kaons, the target dependence is stronger in nucleus-nucleus
collisions than in p-nucleus collisions. 
The $K^+/\pi^+$ ratio is constant for $pA$ collisions but increases
by about $70\%$ in central $SA$ interactions.
 The ratio also increases with centrality.

 A similar effect has been observed by the E802 collaboration for $SiAu$ 
collisions at 14.6GeV/{\it c}, where the $K^+/\pi^+$  ratio is
a factor of $3\pm 1$ larger than for $pp$ collisions of a similar energy
\cite{E80290}. RQMD was able to give a reasonable explanation of this 
result \cite{GON95A} in terms of secondary 
production of $K^+$ by meson-baryon 
interactions, see also \cite{SOR98A}. A similar result was obtained with
ART \cite{ART98A}, while in ARC \cite{ARC92A} the $K^+$ enhancement was caused by
$\Delta \Delta$ interactions.
 RQMD predicts quite well the measured features of the dN/dy of pions and 
kaons,
 but slightly overestimates the $K^+/\pi^+$ ratio in $pA$ collisions.

For $SS$ collisions, isospin conservation implies 
that the total yield  of $K^0_S$ over the full phase
space should equal 
the average of the $K^+$ and $K^-$ yields.
(A similar argument holds for pions).
Comparing our yields with those of NA35 \cite{NA3594A}, we find
 excellent agreement for kaons and reasonable agreement for pions
 \cite{NA3594B}. However the NA35 kaon inverse slopes are significantly larger 
\cite{NA3596A}. The NA36 and WA85 collaborations have measured the 
inverse slopes of charged kaons from $SPb$ and $SW$ collisions respectively 
 and we
are in good agreement with their results \cite{NA3692A,WA8595A}.
Although WA85 does not give absolute yields, their $K^-/K^+$ ratio for
$y=2.3-3.0$ from $SW$ \cite{WA8595A}
 is very close to our value for $SPb$ at $y=2.3-2.9$.

\section{Discussion}
\label{sec-disc}

Fig.~\ref{fg:pislope} shows that the $\pi^+$ inverse 
slopes increase with increasing $p_T$.
Cronin {\it et al}~\cite{cronin}
parametrized their $pA$ pion spectra in terms of $pp$ spectra as follows
\begin{equation}
 \frac{E d^3\sigma}{d^3p}(pA) = A^{\alpha(p_T)}\frac{E d^3\sigma}{d^3p}(pp).
\label{eq:alphapp}
\end{equation}
This worked well for many different nuclei. It was found that $\alpha$
depended only slowly on rapidity but strongly on $p_T$, running from 
0.7 at low $p_T$ to greater than 1 for very high $p_T$. The low $p_T$ 
value of alpha was close to the value 2/3 expected for an opaque nucleus
while $\alpha = 1$ implies full participation of the nucleus.
In \cite{lev83} $\alpha > 1$ was understood in terms of parton scattering. 
The HELIOS collaboration extended Eq.~\ref{eq:alphapp} to nucleus-nucleus
collisions \cite{na3490a} by writing
\begin{equation}
 \frac{E d^3\sigma}{d^3p}(AB) = (AB)^{\alpha(p_T)}\frac{E d^3\sigma}{d^3p}(pp).
\label{eq:alphaab}
\end{equation}
Using Eq.~\ref{eq:alphaab} to compare our $SPb$ and $pBe$ $\pi^+$ data, 
 we find 
that $\alpha = 0.8$ at $p_T=0$ and rises to 1.0 by about 1.2~GeV/$c$. This
is in agreement with recent results on $\pi^0$ production by
WA80 \cite{wa8098a}. HELIOS 
observed  slightly smaller
 values of 
$\alpha$ when comparing negative hadron production from $SW$ and $pW$. This 
may be due to their more backward rapidity range where the density of 
particles is lower.

A prerequisite for creating a quark gluon plasma is to produce a hot dense
nuclear system in heavy ion collisions. Such a system would be likely to 
undergo transverse expansion which would affect the inverse slopes of the 
particle spectra. The effect of such a velocity boost would be more pronounced
on heavier particles. 
Fig.~\ref{fg:tpikp} shows that the difference in inverse slopes between 
pions, kaons and protons increases with system size.
This suggests that the effect of transverse flow becomes stronger for
larger systems.
NA44 has previously reported an increase with system size
in the inverse slopes of kaons, 
protons and antiprotons produced at mid-rapidity in symmetric collision
systems \cite{NA44pbpb}. These data extend this trend to asymmetric systems.
Since the effect of the boost increases 
with mass, the inverse slopes of pions do not increase much with
system size. 

 The target dependence of dN/dy 
is stronger in nucleus-nucleus
collisions than in p-nucleus collisions probably because 
additional particles are made in secondary collisions of the produced 
particles.
The increase of the $K^+/\pi^+$ ratio with system size suggests that the
fraction of strange quarks is increasing. 
In RQMD, rescattering increases kaon production by $50\%$ for $SS$,
 mainly because
of associated production of $K$ and $\Lambda$ in 
meson-baryon collisions \cite{Sor92a}. However, the total multiplicity 
remains essentially the same if rescattering is turned on or off in the code.

\begin{figure}[hb]
  \vspace{-1.0cm}
 \begin{center}
    \mbox{
     \epsfxsize=12cm
     \epsffile{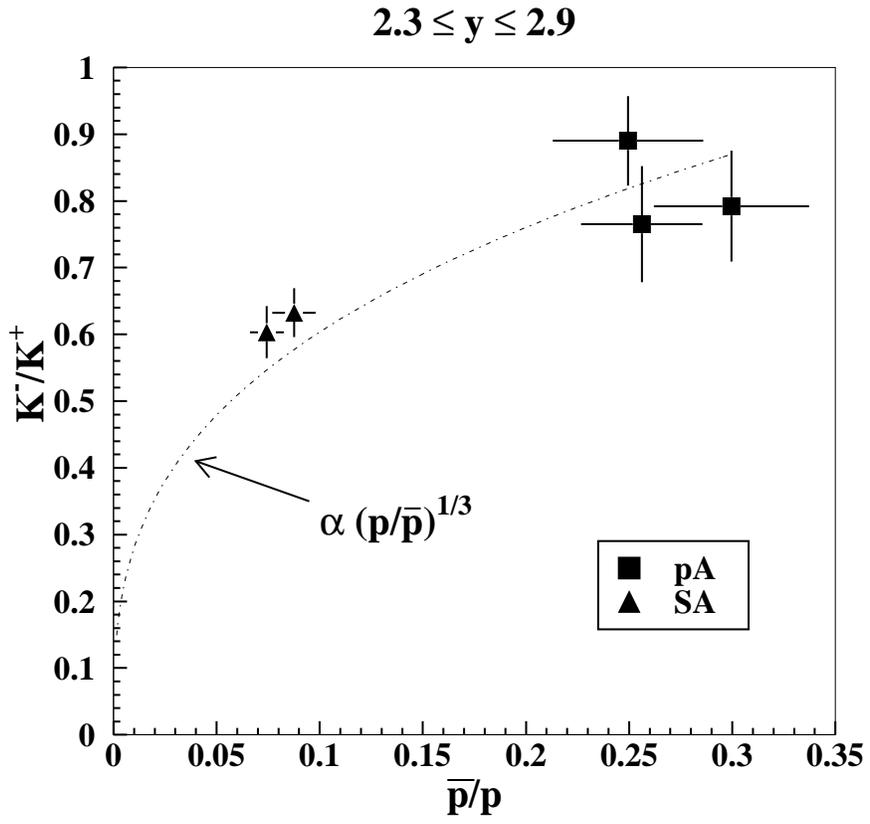}
      }
  \end{center}
  \caption{$K^-/K^+$ versus $\bar{p}/p$ for various systems at $y=2.3-2.9$.}
 \label{fg:kmpvpmp}
\end{figure}

Fig.  \ref{fg:kmpvpmp} shows the 
  $K^-/K^+$ ratio versus our $\bar{p}/p$ ratio \cite{NA44pmp} 
 for various systems at y=2.3-2.9. 
If we have a system of quarks in chemical and thermal equilibrium, then 
ratios of particles and antiparticles
 can be described by ratios of the chemical potential of
the  constituent quarks to the temperature \cite{KOCH83A}.
 Since the quark content of
the proton is $uud$ and $K^+ = u\bar{s}$ ($K^-= \bar{u}s$), then
 $K^-/K^+ = e^{2\mu_s/T} (\bar{p}/p)^{1/3}$
where $\mu_s$ is the chemical potential of strange quarks and $T$ is the
temperature.
 If $\mu_s/T$ changes slowly from $pBe$ to $SPb$ then
we would expect $K^-/K^+ \propto (\bar{p}/p)^{1/3}$. The data are close to this
simple form.
As the colliding system gets larger, the baryon density increases reducing the
 fraction of ${\bar u}$ quarks and  driving down the $K^-/K^+$ ratio.

\begin{figure}[hb]
  \vspace{-1.0cm}
 \begin{center}
    \mbox{
     \epsfxsize=12cm
     \epsffile{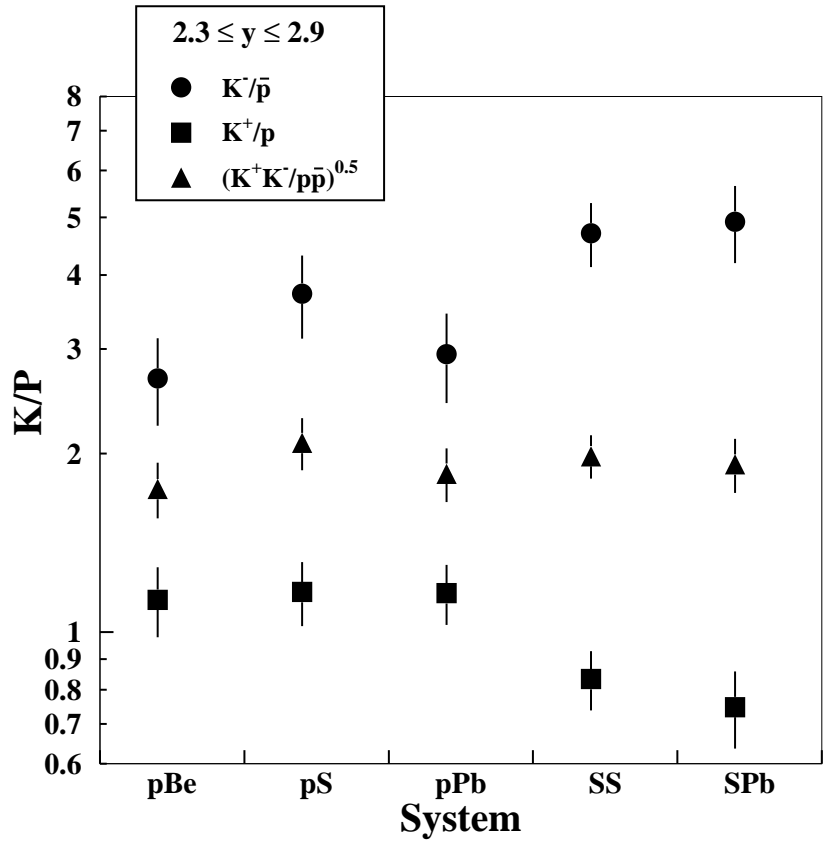}
      }
  \end{center}
  \caption{The $K^+/p$ and $K^-/\bar{p}$  ratios versus system.}
 \label{fg:PKPBARK}
\end{figure}

Fig.~\ref{fg:PKPBARK} shows 
the ratios $K^+/p$ , $K^-/\bar{p}$ and their geometric mean
$\sqrt{\frac{K^+\cdot K^-}{p \cdot \bar{p}}} $ versus system. 
The $K^+/p$ ratio decreases while $K^-/\bar{p}$ increases and 
$\sqrt{\frac{K^+\cdot K^-}{p \cdot \bar{p}}} $ stays constant.
Again assuming chemical and thermal equilibrium, 
this implies that 
the number of strange quarks is proportional to the 
product of the numbers of u and d quarks. Such behavior is consistent with the
increase of the $K^+/\pi^+$ ratio shown in Fig.~\ref{fg:kpidndy}.
This suggests that in $SA$ collisions strange quarks
are still suppressed somewhat with respect to u and d quarks. In 
thermal language, the chemical potential of the u and d quarks is not yet
equal to the mass of the strange quark.

\section{Conclusions}

Several features of the data suggest a strong build up of rescattering
and hence energy density
as we collide larger ions. These are:
\begin{itemize}
\item The stronger target dependence of yields and slopes for
$SA$ collisions than $pA$ collisions;
\item The increase of the pion inverse slopes with $m_T$;
\item The increase in the $K^+/\pi^+$ ratio with system size.
\end{itemize}
If particles undergo many rescatterings, 
pressure builds up and this causes radial flow. The increase of the 
inverse slopes with
particle mass and the increase of the difference in slopes for larger 
systems supports the existence of radial flow. However the extraction
of a flow velocity is probably best done from a combined analysis of our
interferometry measurements and single particle spectra.

At central rapidity the $K^-/K^+$ ratio falls as the baryon density 
increases
because of the increasing scarcity of the anti-up quarks required to form
$K^-$. The increasing $K^+/\pi^+$ ratio as well as the constancy of 
$\sqrt{\frac{K^+\cdot K^-}{p \cdot \bar{p}}} $ implies that the fraction
of strange quarks in the hadronic system is increasing. 

\section{Acknowledgements}
NA44 is grateful to the staff of the CERN PS--SPS
accelerator complex for their excellent work. We thank the technical
staff at CERN and the collaborating institutes for their valuable
contributions. We are also grateful for the support given by the
 Austrian Fonds zur F\"orderung der Wissenschaftlichen Forschung (grant
P09586); the Science Research Council of Denmark; the Japanese Society
for the Promotion of Science; the Ministry of Education, Science and
Culture, Japan; the Science Research Council of Sweden; the US
W.M.~Keck Foundation; the US National Science Foundation; and the US
Department of Energy.

\begin{table}[tb]
 \begin{center}\mbox{ }
  \caption{Fraction of events satisfying the interaction trigger,
 target thickness and 
number of events at each setting for each particle type. The target thickness, $\lambda$, is quoted in
nuclear collision lengths for the given system.}
  \label{tb:data}
   \begin{tabular}{|l|c|c|c|r|r|r|r|r|r|} \hline
 &  {\bf Event } & {\bf Angle}  &
 {\bf $\lambda$}  &
 \multicolumn{2}{c|}{\bf 4GeV/$c$} &
 \multicolumn{2}{c|}{\bf 8GeV/$c$} & {\bf 4GeV/$c$} & {\bf 8GeV/$c$}\\
 & {\bf Fraction} & mrad & \% & 
{\bf $K^+$}  & {\bf $K^-$}  & {\bf $K^+$}  & {\bf $K^-$} &\multicolumn{1}{c|}{\bf $\pi^+$}&\multicolumn{1}{c|}{\bf $\pi^+$} \\ \hline
$pBe$ & 84 $\pm$ 1.5\% &  44 & 3.3   &  22723 &  47498 &   3104 &  11594 &  15583 &  69088 \\ \cline{3-10}
      &                & 131 & 3.3   &   5801 &   7948 &   2673 &   2671 &  38240 &  15677 \\ \hline
$pS $ & 90 $\pm$ 2  \% &  44 & 3.3   &  40389 &  39161 &   1657 &  11263 &  15594 &  67043 \\ \cline{3-10}
      &                & 131 & 3.3   &  22949 &   4661 &      0 &      0 &      0 &      0 \\ \hline
$pPb$ & 97 $\pm$ 3  \% &  44 & 4.7   &   9591 &  45303 &   1304 &  12268 &  13147 &  22686 \\ \cline{3-10}
      &                & 131 & 9.9   &   8898 &   1304 &   4674 &   3497 &  25553 &   5195 \\ \hline
$SS $ & 8.7$\pm$  .5\% &  44 & 6.6   &   9475 &  11875 &   6539 &  10236 &  20201 &  21155 \\ \cline{3-10}
      &                & 131 & 6.6   &   9907 &   9996 &   2780 &      0 &  25898 &  16777 \\ \hline
$SPb$ &10.7$\pm$  .6\% &  44 & 5.9   &   8893 &   8695 &      0 &  14034 &   3668 &   4613 \\ \cline{3-10}
      &                & 131 & 5.9   &  18653 &  20028 &   3645 &      0 &  22667 &  15949 \\ \hline
   \end{tabular}
  \end{center}
\end{table}
\begin{table}[t]
  \caption{Systematic errors on the inverse slopes and dN/dy.}
  \label{tb:errors}
 \begin{center}\mbox{ }
   \begin{tabular}{|c|c|c|c|c|c|c|} \hline
             &  \multicolumn{3}{c|}{\bf Kaons} &  \multicolumn{3}{c|}{\bf Pions} \\
{\bf System} &  \multicolumn{2}{c|}{\bf Error on Inverse Slope} & {\bf Error} 
             &  \multicolumn{2}{c|}{\bf Error on Inverse Slope} & {\bf Error} \\
  &  {\bf y=2.3-2.9} & {\bf y=2.4-3.5} & {\bf on dN/dy} &  {\bf y=2.4-3.0} & {\bf y=3.1-4.0} &  {\bf on dN/dy} \\ \hline
$pBe$ & 10 MeV/{\it c} & 10 MeV/{\it c} & 9\%  & 10 MeV/{\it c} & 10 MeV/{\it c} & 9\% \\ 
$pS$  & 10 MeV/{\it c} & 10 MeV/{\it c} & 9\%  & 10 MeV/{\it c} & 10 MeV/{\it c} & 9\% \\ 
$pPb$ & 10 MeV/{\it c} & 10 MeV/{\it c} & 10\% & 10 MeV/{\it c} & 10 MeV/{\it c} & 10\% \\
$SS$  & 20 MeV/{\it c} & 10 MeV/{\it c} & 9\%  & 10 MeV/{\it c} & 10 MeV/{\it c} & 8\% \\ 
$SPb$ & 20 MeV/{\it c} & 10 MeV/{\it c} & 14\% & 10 MeV/{\it c} & 10 MeV/{\it c} & 9\% \\ \hline
   \end{tabular}
  \end{center}
\end{table}
\begin{table}[t]
 \begin{center}\mbox{ }
  \caption{Inverse slopes (in MeV/{\it c}) for $K^+$, $K^-$ and $\pi^+$ with
 statistical errors.}
  \label{tb:slope}
   \begin{tabular}{|l|c|c|c|c|c|c|} \hline
 & \multicolumn{2}{c|}{\bf y=2.3-2.9} &
   \multicolumn{2}{c|}{\bf y=2.4-3.5} & {\bf y=3.1-4.0} & {\bf y=2.4-3.0}\\
 & \multicolumn{2}{c|}{\bf $k_T\le 0.33$ GeV/{\it c}} &
   \multicolumn{2}{c|}{\bf $k_T\le 0.84$ GeV/{\it c}} & 
   {\bf $k_T\le 0.64$ GeV/{\it c}}& {\bf $k_T=0.3-1.2$ GeV/{\it c}} \\
 & {\bf $K^+$}   &{\bf $K^-$}&{\bf $K^+$}&{\bf $K^-$}& {\bf $\pi^+$}& {\bf $\pi^+$}\\ \hline
$pBe$&$138\pm 7$ &$145\pm 8$ &$154\pm 8$&$153\pm 4$& $148\pm3$ & $169\pm3$ \\
$pS $&$152\pm 4$ &$154\pm10$ &$163\pm14$&$160\pm12$& $139\pm3$ &            \\
$pPb$&$151\pm 6$ &$147\pm18$ &$172\pm 9$&$152\pm 5$& $145\pm3$ & $167\pm2$ \\
$SS $&$181\pm 7$ &$177\pm 6$ &$159\pm 5$&$163\pm 5$& $154\pm5$ & $179\pm4$ \\
$SPb$&$197\pm13$ &$213\pm10$ &$181\pm 8$&$175\pm 4$& $165\pm9$ & $192\pm5$ \\ \hline
   \end{tabular}
  \end{center}
\end{table}
\begin{table}[t]
 \begin{center}\mbox{ }
  \caption{dN/dy for $K^+$, $K^-$ and $\pi^+$ with statistical and systematic errors.}
  \label{tb:dndy}
   \begin{tabular}{|l|c|c|c|c|c|} \hline
 & \multicolumn{2}{c|}{\bf y=2.3-2.9} & \multicolumn{2}{c|}{\bf y=2.4-3.5} & {\bf y=2.4-4.0} \\
 & {\bf $K^+$}  & {\bf $K^-$}         & {\bf $K^+$}  & {\bf $K^-$}  & {\bf $\pi^+$}\\
$pBe$&$ .18\pm .01\pm .02$&$ .14\pm .01\pm .01$&$ .15\pm .02\pm .02$&$ .13\pm .01\pm .01$&$2.1\pm 0.2\pm 0.2$ \\
$pS $&$ .24\pm .01\pm .02$&$ .21\pm .01\pm .02$&$ .24\pm .03\pm .02$&$ .18\pm .02\pm .02$&$2.4\pm 0.2\pm 0.2$\\
$pPb$&$ .29\pm .01\pm .03$&$ .22\pm .02\pm .02$&$ .27\pm .02\pm .03$&$ .21\pm .02\pm .02$&$3.2\pm 0.2\pm 0.3$\\
$SS $&$ 3.76\pm .19 \pm .35 $&$ 2.4 \pm .1  \pm .2  $&$ 4.9 \pm .4  \pm .5 $&$ 3.4 \pm .3 \pm .3 $ &$29.7\pm 1.8 \pm 2.2$\\
$SPb$&$ 9.0 \pm .5  \pm 1.2 $&$ 5.4 \pm .2  \pm .7  $&$13.1 \pm 1.6 \pm 1.8$&$ 6.7 \pm .4 \pm .9 $&$62  \pm 7   \pm 5  $\\ \hline
   \end{tabular}
 \end{center}
\end{table}



\end{document}